\documentclass[12pt]{article}
\usepackage{epsfig}
\usepackage{subfigure}

\begin{document}
\mbox{ }\\[-1cm]
\mbox{ }\hfill \\
\mbox{ }\hfill \\

\vskip 1cm

\begin{center}
{\Large\bf Investigation of the transverse beam dynamics in \\
           the thermal wave model with a functional method}\\[1cm]   
Ji-ho Jang\footnote{jangjh@kaeri.re.kr}, Yong-sub Cho, Hyeok-jung Kwon
\end{center}

\bigskip 

\begin{center}
{\it Korea Atomic Energy Research Institute, Daejeon 305-353, Korea}
\end{center}

\vskip 2.5cm

\begin{abstract}
We investigated the transverse beam dynamics in a thermal wave
model by using a functional method. It can describe the beam
optical elements separately with a kernel for a component. The method can be
applied to general quadrupole magnets beyond a thin lens approximation 
as well as drift spaces.
We found that the model can successfully describe the PARMILA simulation result 
through an FODO lattice structure for the Gaussian input beam 
without space charge effects.
\vskip 0.5cm

\noindent
PACS number(s): 29.27.-a, 29.27.Eg
\end{abstract}
\newpage

The thermal wave model is an efficient way to study the beam dynamics of 
relativistic charged particles. The Schr\"{o}dinger-type equation in the 
model governs the time evolution of the beam wave function whose squared 
magnitude is proportional to the particle number densities\cite{FM:1991}.
The model has successfully explained the filamentation of a particle beam 
and the
self-pinching equilibrium in collisionless plasma\cite{FS:1992}.
It was also used to estimate the luminosity in a linear collider where
a spherical aberration was present\cite{FM:1992}.
The model can also provide some insight into a halo formation by introducing 
a Gaussian slit\cite{KP:2000}.

Transverse beam dynamics in a one spatial dimension 
is another application area of the thermal wave model.
In Ref. \cite{FGM:1994}, the authors investigated the beam wave function through 
a quadrupole magnet with sextupole and octupole perturbations followed by a
long drift space under a thin lens approximation.
There is also a paper on the phase space behavior of particle beams in the
transverse directions
where the Wigner and Husimi functions are used as the phase space distribution
functions\cite{FGMM:1995}.

In this work, we investigate the transverse beam dynamics in a two dimensional
trace ($x-x'$ or $y-y'$) space in the thermal wave 
model by using the functional integral method \cite{H:1992}.
Because the method can be extended to general lattice structures including
quadrupole magnets beyond a thin lens approximation limit and
it can treat the beam optical elements individually, it is possible 
to systematically analyze a beam motion in a realistic
environment such as an FODO lattice. 
We found that the model can successfully explain the PARMILA\cite{P:2002} 
simulation results with Gaussian input beams in a two dimensional trace 
space under the condition that the space charge effects are negligible. 
We note that this method can explain a low energy particle behavior 
as well as the relativistic motion of the charged particles
if the important interactions are related to 
the external linear optical elements such as 
the quadrupole magnets and the random motion described by a beam emittance.

In the thermal wave model, the time evolution of the beam wave function for 
the relativistic charged 
particles can be described by the Schr\"{o}dinger-type equation in the
transverse directions. Because the beam dynamics are usually described in the 
two dimensional trace ($x-x'$ or $y-y'$) space, 
it is important to see whether the one dimensional version of the equation 
can explain the beam dynamics in the projected space or not. 
The one dimensional Schr\"{o}dinger-type equation in $x$ direction is given by 
\begin{eqnarray}
\label{eq:equation_x}
i~\epsilon_x~\frac{\partial~\psi_x(x,z)}{\partial z} =
-~\frac{\epsilon_x^2}{2}~\frac{\partial^2~\psi_x(x,z)}{\partial x^2}
+ U_x(x,z)~\psi_x(x,z),
\end{eqnarray}
where $z = ct$ is the longitudinal distance of the beam movement and
$U_x(x,z)\equiv u_x(x,z)/m_0 \gamma_r \beta_r^2 c^2$ is the dimensionless 
potential with the relativistic parameters, $\beta_r = v/c$ and 
$\gamma_r =(1-\beta_r^2)^{-1/2}$.
The parameter $\epsilon_x$ is related to the emittance of the particle
distribution in the $x-x'$ space, which is explained later.
The transverse particle distribution can be obtained by the squared magnitude 
of the beam wave function, $N \left| \psi_x(x,z) \right|^2$ with the particle
number of $N$.
In this convention, 
the beam wave function satisfies the normalization condition as follows,
$ \int_{-\infty}^{\infty} \left| \psi_x(x,z) \right|^2 dx = 1 $.
The corresponding equation for the time evolution of the wave function 
in the $y$ direction can be obtained by replacing $x$ with $y$ in
Eq.~(\ref{eq:equation_x}). 
We note that the time evolution of the beam wave function in $y$ direction is
independent of that in the $x$ direction because the one dimensional
Schr\"{o}dinger-type equation in the $y$ direction includes the different 
parameter of $\epsilon_y$ and we considered the linear external forces only.
In the following analysis, the parameters and the functions in the $x$ direction 
are used without subscript if there is no confusion.

We can solve the differential equation by imposing the following two boundary
conditions, $\sigma^2(z=0) = \sigma_0^2$ and 
$\left.\frac{1}{\sigma} \frac{d \sigma}{d z} \right|_{z=0} = \frac{1}{\rho_0}$
\cite{FGM:1994}.
The $\sigma$ denotes the root mean square (rms) size of the beam distribution
and $\rho$ is the curvature radius of the beam wave function along the beam
direction.

Another efficient way to solve the differential equation is known as
the functional integral method \cite{H:1992}
where the resulting wave function is given by the product 
of a kernel (or propagator) and the initial beam wave function,
\begin{eqnarray}
\label{eq:formal_solution}
\psi(x_f,z_f) =  
\int_{-\infty}^{\infty} dx_i K(x_f,z_f ; x_i,z_i)~\psi(x_i,z_i).
\end{eqnarray}

Since a kernel represents an optical element like a quadrupole magnet or 
a drift space, the functional method can separate a multi-components problem 
into several single-component problems. 
This property is the main advantage of this
functional method in the thermal wave model.

We can obtain the kernels from the path integral method \cite{H:1992}
directly as follows,
\begin{eqnarray}
K(x_f,z_f ; x_i,z_i) = \int {\mathcal D}[ x(z) ]~ e^{i S(z)/\epsilon},
\end{eqnarray}
where $S(z)=\int_{z_i}^{z_f} dz {\cal L}(x(z),x'(z))$ is called the action. 
The Lagrangian, ${\cal L}$, of a system is the difference of the kinetic and 
potential energy terms.

In this work, we will restrict our attention to a system consisting of
quadrupole magnets and drift spaces.
The potential energy terms of the beam optical elements are given by
\begin{eqnarray}
U(x)=\left\{
\begin{tabular}{lll}
0                   &~~~~~&\mbox{for a drift space}, \\
$\frac{1}{2} k_1 x^2$ &~~~~~&\mbox{for a focusing quadrupole magnet},
\end{tabular}
\right.
\end{eqnarray}
where $k_1$ is positive in the focusing case.
The potential term for the defocusing magnet is $-(k_1/2) x^2$.


The kernel, $K_0$, for a drift space which has no potential term is given by
\begin{eqnarray}
K_0 (x_f,z_f;x_i,z_i) &=& \left( \frac{1}{2 \pi i \epsilon (z_f - z_i)}\right)^{1/2}
e^{\frac{i}{2 \epsilon (z_f - z_i)} (x_f-x_i)^2}.
\end{eqnarray}

The kernel, $K_f$, for the focusing quadrupole magnet is given by
\begin{eqnarray}
\label{eq:kernel_focusing_quad}
K_f (x_f,z_f;x_i,z_i) &=& \left( \frac{\sqrt{k_1}} 
   {2 \pi i \epsilon \sin (\sqrt{k_1}(z_f-z_i))} \right)^{1/2} 
   e^{i \frac{\sqrt{k_1}}{2 \epsilon} \left[(x_f^2+x_i^2) \cot \sqrt{k_1}z -
   2 x_f x_i \csc \sqrt{k_1} z \right]}.
\end{eqnarray}
For the defocusing case, the kernel is obtained easily by replacing 
the cot and csc functions in Eq.(\ref{eq:kernel_focusing_quad}) 
with coth and csch functions, respectively.

Since the potential energy terms are related to linear forces only,
the integration in Eq. (\ref{eq:formal_solution}) 
becomes very simple if the initial beam wave
function is a Gaussian-type such as
\begin{eqnarray}
\label{eq:input_wave_function}
\psi_1(x,0)=\left( \frac{1}{2 \pi \sigma_1^2} \right)^\frac{1}{4}
          \exp \left[-\frac{x^2}{4 \sigma_1^2} + 
          i \left( \frac{x^2}{2 \epsilon \rho_1} + \theta_1 \right) \right],
\end{eqnarray}
where $\sigma_1$, $\rho_1$, $\theta_1$ are the initial values of the rms beam
size, the curvature radius, and the input phase, respectively.

After the input beam passes through a linear optical element, 
the beam wave function remains the Gaussian-type such as
\begin{eqnarray}
\label{eq:solution}
\psi_2(x,z)
= \left( \frac{1}{2 \pi \sigma_2^2(z)} \right)^\frac{1}{4}
          \exp \left[-\frac{x^2}{4 \sigma^2_2(z)} + 
          i\left( \frac{x^2}{2 \epsilon \rho_2(z)} +\theta_1+\theta_2(z) \right)
          \right].
\end{eqnarray}
The different forms of the parameter functions, $\theta_2(z), \sigma_2(z)$, and 
$\rho_2(z)$, characterize the properties of each optical element.

In a drift space, the functions are given by
\begin{eqnarray}
\sigma^2_2(z) &=& \sigma^2_1 \left[ \left( \frac{\epsilon z}{2 \sigma^2_1} 
               \right)^2 + \left( 1+ \frac{z}{\rho_1} \right)^2 \right],
\\
\tan 2\theta_2(z) &=& -\frac{\epsilon}{2 \sigma_1^2} \frac{z \rho_1}{z+\rho_1}, \\
\frac{1}{\rho_2(z)} &=& \frac{1}{\rho_1}
    \left[ \frac{\rho_1}{z} - \left( \frac{\sigma_1}{\sigma_2(z)} \right)^2
    \left(1+\frac{\rho_1}{z} \right) \right].
\end{eqnarray}

In a focusing quadrupole magnet, they are given by
\begin{eqnarray}
\sigma^2_2(z) &=& \sigma^2_1 \left[ \left( \cos(\sqrt{k_1}z) + 
          \frac{1}{\sqrt{k_1} \rho_1} \sin(\sqrt{k_1}z) \right)^2 +
          \left( \frac{\sigma_0}{\sigma_1} \right)^4 \sin^2(\sqrt{k_1} z) \right],
\\
\tan 2\theta_2(z) &=&-\frac{\left( \frac{\sigma_0}{\sigma_1} \right)^2
                    \sin (\sqrt{k_1} z)}{\cos (\sqrt{k_1} z)+ 
                     \frac{1}{\sqrt{k_1}\rho_1}\sin (\sqrt{k_1} z)},\\
\frac{1}{\rho_2(z)} &=& \frac{1}{\rho_1} \left( \frac{\sigma_1}{\sigma_2(z)} \right)^2
    \left[ \cos(2\sqrt{k_1}z) + \frac{1}{2} \left\{ 
    \frac{1}{\sqrt{k_1}\rho_1}+ \sqrt{k_1}\rho_1
    \left( \left( \frac{\sigma_0}{\sigma_1} \right)^4 -1 \right) \right\}
    \sin(2 \sqrt{k_1}z)
    \right],
\nonumber \\
\end{eqnarray}
with $\sigma_0^2 = \epsilon / ( 2 \sqrt{k_1} )$.
For a defocusing lens, the functions can be
obtained by replacing $\sqrt{k_1}$ for the focusing case with $i \sqrt{k_1}$.
We can easily check to see if Eq.~(\ref{eq:solution}) is the solution of
Eq.~(\ref{eq:equation_x}) by inserting the obtained beam wave function into the 
differential equation.

First of all, we studied how to relate the model parameters, 
$\sigma_1, \rho_1, \epsilon$, of the input Gaussian wave function in 
Eq. (\ref{eq:input_wave_function}) to the twiss parameters and the unnormalized
rms emittance, $\alpha_1, \beta_1, \epsilon_{rms}$.
Because $\beta$ is defined as $\sigma^2/\epsilon_{rms}$ 
for the Gaussian distribution, we obtained 
$\sigma_1=\sqrt{\epsilon_{rms} \beta_1}$.
From the definition of $1/\rho \equiv (1/\sigma)(d \sigma / d z)$ 
\cite{FGM:1994}, 
we can easily obtain $1/\rho_1 = - \alpha_1/\beta_1$ where we used 
$ d \beta/ d z = -2 \alpha$.
Motivated by the quantum mechanical relation between the wave functions in the
configuration and momentum spaces, we defined the wave function in 
the $x'$ space as the Fourier transformation of the Gaussian beam wave function 
as follows,
\begin{eqnarray}
\label{eq:phi_x}
\phi_1(x') &\equiv& 
             \frac{1}{\sqrt{2 \pi \epsilon}} \int_{-\infty}^{\infty} dx
             \exp \left[ -i \frac{x x'}{\epsilon}  \right] \psi_1(x,0) \nonumber \\
         &=& \left( \frac{1}{2 \pi \sigma_{1d}} \right)
             \exp \left[ -\frac{(x')^2}{4 \sigma_{1d}^2} + 
             i \left\{ - \frac{(x')^2}{2 \epsilon \rho_{1d}} 
             + \theta_1 + \theta_{1d} \right\} \right]
\end{eqnarray}
where 
\begin{eqnarray}
\label{eq:sigam1d}
\sigma_{1d}^2 &=& \sigma_1^2 \left[ 
                  \left( \frac{\epsilon}{2 \sigma_1^2} \right)^2 +
                  \frac{1}{\rho_1^2} \right], \\
\label{eq:rho1d}
\rho_{1d} &=& \rho_1 \left[ \left( \frac{\epsilon}{2 \sigma_1^2} \right)^2
              + \frac{1}{\rho_1^2} \right],
\end{eqnarray}
with $\tan (2 \theta_{1d}) = 2 \sigma_1^2/(\epsilon \rho_1)$. 
The initial particle 
distribution in the $x'$ space is proportional to $|\phi_1(x')|^2$.
Because we can define $\gamma$ as $\sigma_{d}^2 / \epsilon_{rms}$ for a Gaussian
distribution in $x'$ space, we obtain
$\sigma_1^2 \sigma_{1d}^2 = \epsilon_{rms}^2 \beta_1 \gamma_1$.
Comparing it with 
$\sigma_1^2 \sigma_{1d}^2 = \epsilon^2/4 + \epsilon_{rms} \alpha_1^2$ 
which can be obtained from Eq. (\ref{eq:sigam1d}), 
we can obtain $\epsilon = 2 \epsilon_{rms}$ where we used the relation between
twiss parameters, $\beta_1 \gamma_1 - \alpha_1^2 =1$. 
We also obtained $1/\rho_{1d} = - \alpha_1/\gamma_1$ from 
Eq. (\ref{eq:rho1d}). 
We note that the relations between the model and physical parameters are 
valid for the wave functions at each of the beam optical elements.

We note that the above analysis for the time evolution of the beam in 
$x-x'$ space is also valid in the $y-y'$ space if we use 
the one-dimensional Schr\"{o}dinger-type equation in $y$ direction with
the emittance parameter of $\epsilon_y$. 
In the following analysis, we studied the time evolution of the beam wave 
functions in both the horizontal ($x-x'$) and vertical ($y-y'$) spaces 
by using the equations with different emittance parameters, 
$\epsilon_x$ and $\epsilon_y$.

\begin{figure}[htb]
 \centering
 \hspace{-1cm}
  \subfigure[]{
  \includegraphics[angle=0, scale=0.7]{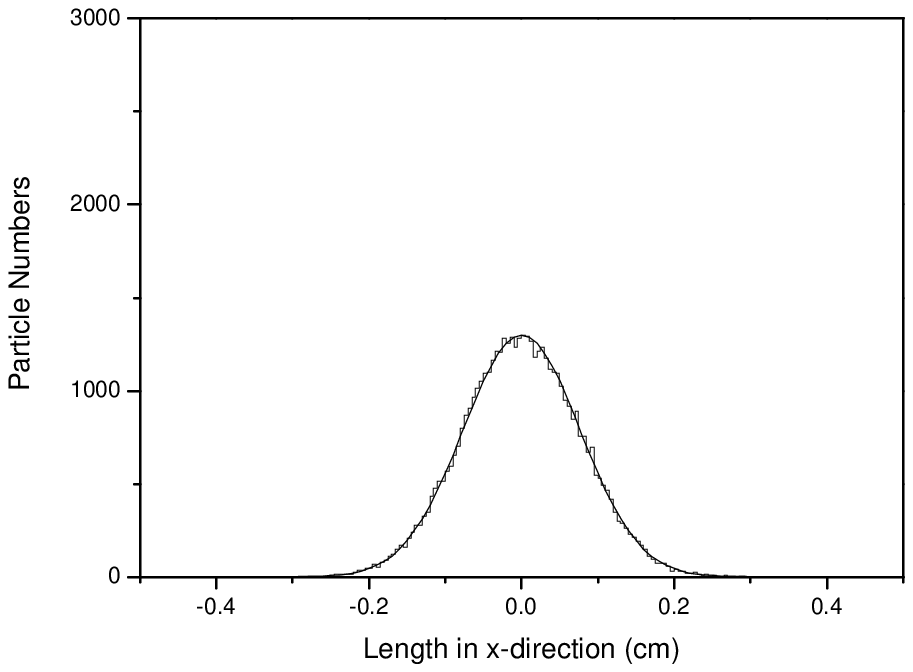}
  \label{fig:input_beam_x}
  }\quad
  \subfigure[]{
  \includegraphics[angle=0, scale=0.7]{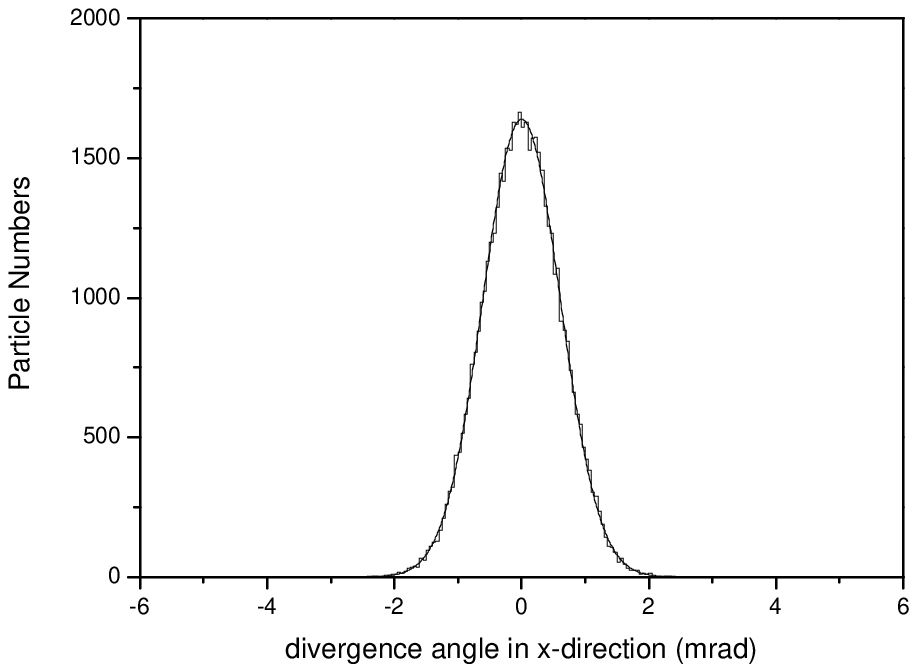}
  \label{fig:input_beam_xp}
  }\quad
 \caption{\label{fig:input_beam_h}
          Particle distributions of the input beam in (a) $x$-direction and 
          (b) $x'$-direction of the horizontal trace space. 
          The histograms and real lines represent the 
          PARMILA results and the model predictions, respectively.}
 \end{figure}

\begin{figure}[htb]
 \centering
 \hspace{-1cm}
  \subfigure[]{
  \includegraphics[angle=0, scale=0.7]{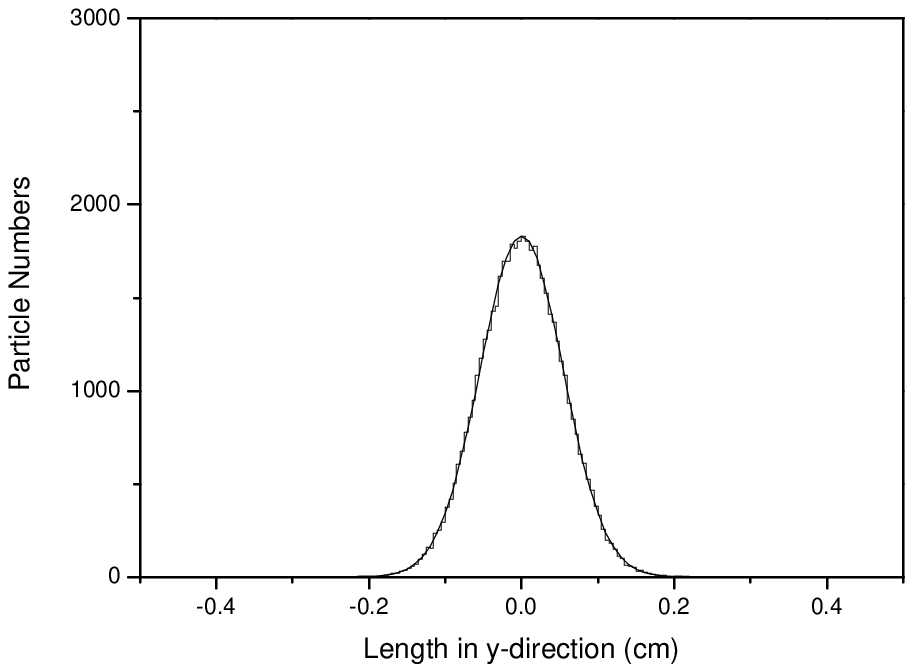}
  \label{fig:input_beam_y}
  }\quad
  \subfigure[]{
  \includegraphics[angle=0, scale=0.7]{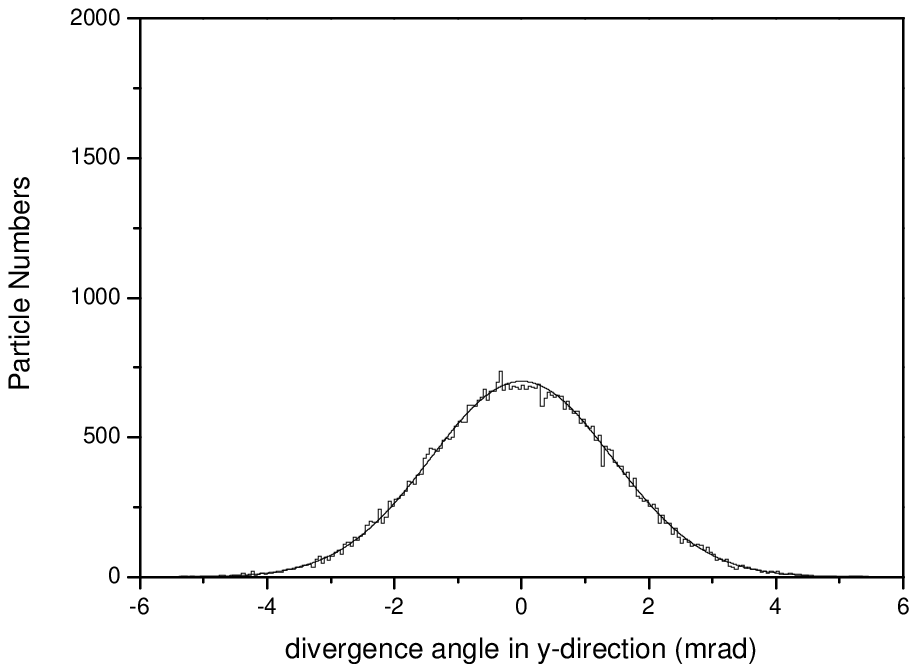}
  \label{fig:input_beam_yp}
  }\quad
 \caption{\label{fig:input_beam_v}
          Particle distributions of the input beam in (a) $y$-direction and 
          (b) $y'$-direction of the vertical trace space. 
          The histograms and real lines represent the 
          PARMILA results and the model predictions, respectively.}
 \end{figure}

In order to check on the validity of the solutions, we compared them with
the PARMILA simulation results with 50,000 macro particles through the
FODO lattice in the horizontal direction. It corresponds to the DOFO lattice
in the vertical direction.
The field gradient and effective length of the quadrupole magnets 
in the lattice are 10.0 T/m and 0.2 m, respectively. 
The length of the drift spaces is 0.5 m. 
The particle type is proton with a kinetic energy of 100 MeV. 
We selected a random distribution of the particles in the trace spaces 
and neglected the space charge effects. 
The weighting function of the distribution is a Gaussian-type truncated 
at four times the standard deviation.
Figure~\ref{fig:input_beam_x} (\ref{fig:input_beam_y}) and
Figure~\ref{fig:input_beam_xp} (\ref{fig:input_beam_yp}) show the 
particle distributions of the input beam in the $x(y)$ and $x'(y')$ directions 
of the horizontal (vertical) trace space, respectively.
The histograms are the PARMILA result with 50,000 macro particles. 
The real lines represent the Gaussian input beam for the model calculation.
They are obtained by fitting the histograms of the PARMILA results.
In the all figures of this work, we used the same normalization factors of the
distribution functions as those of the input functions.
We found that the beam wave function of Eq. (\ref{eq:phi_x}) describes the 
initial particle distribution very well in both $x'$ and $y'$ directions.

The properties of the input beam are summarized in Table~\ref{table:input_beam}.
From the relations between the model and physical parameters,
we can obtain the input values of the model parameters as follows,
\begin{center}
\begin{tabular}{lll}
$\sigma_{1x}= 0.76$ mm  & $\rho_{1x}=-1.48$ m  & for the horizontal direction, \\
$\sigma_{1y}= 0.55$ mm  & $\rho_{1y}=-0.40$ m & for the vertical direction, \\
$k_1 = 6.74~\mbox{m}^{-2}$, &                    &
\end{tabular}
\end{center}
where $k_1 = q G/(m \gamma_r \beta_r c)$ with the quadrupole field gradient,
$G$.

\begin{figure}[htb]
 \centering
 \hspace{-1cm}
  \subfigure[]{
  \includegraphics[angle=0, scale=0.6]{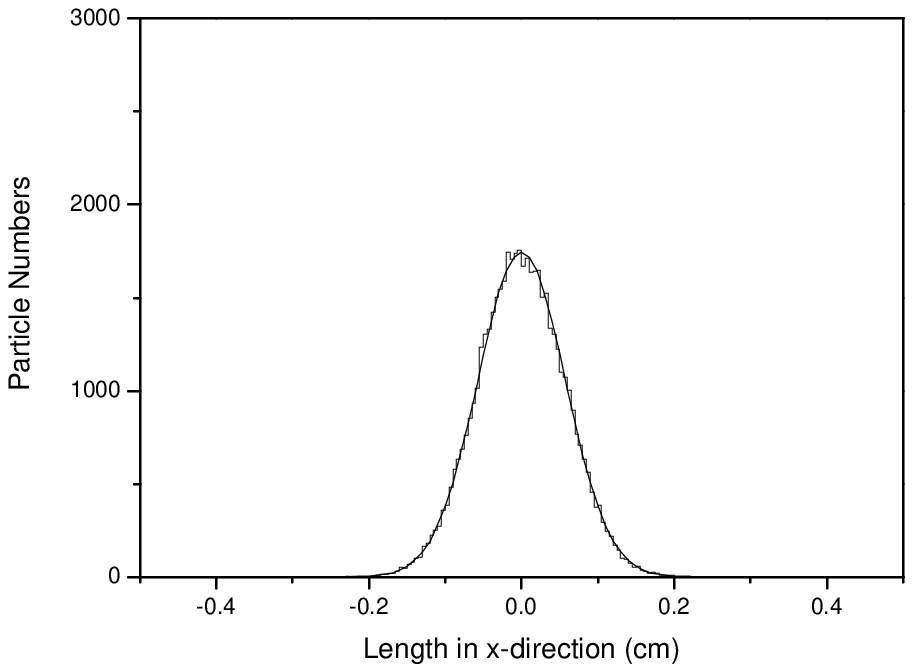}
  \label{fig:horizontal_beam_f}
  }\quad
  \subfigure[]{
  \includegraphics[angle=0, scale=0.6]{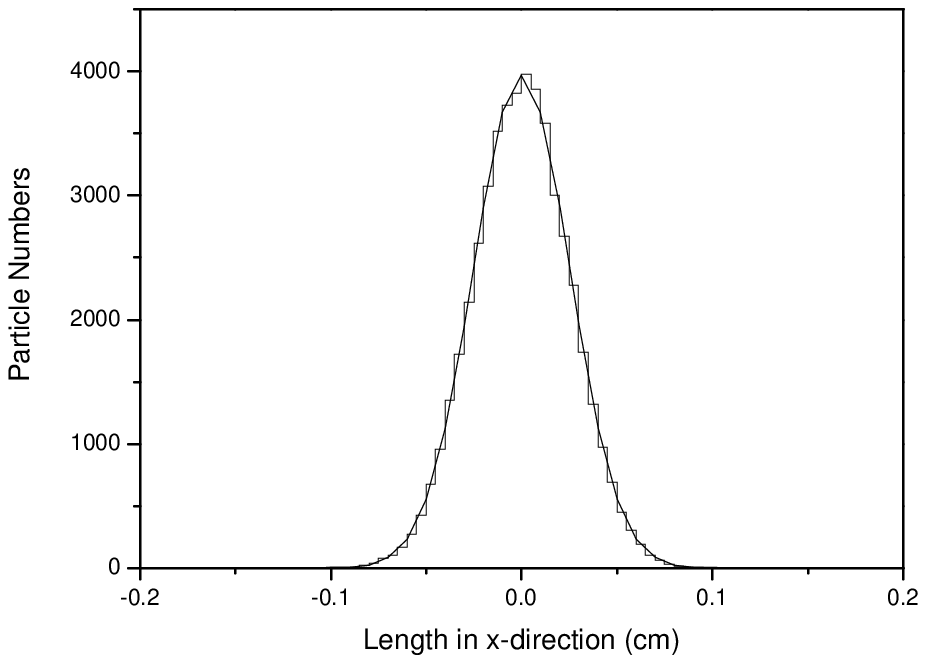}
  \label{fig:horizontal_beam_o}
   }\\
 \hspace{-1cm}
    \subfigure[]{
  \includegraphics[angle=0, scale=0.6]{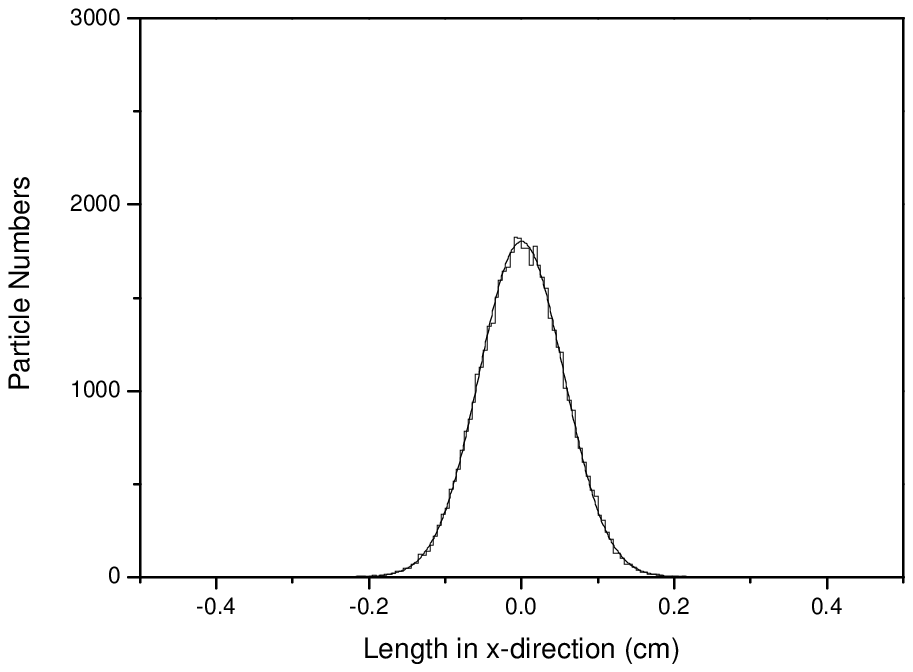}
  \label{fig:horizontal_beam_d}
  }\quad
  \subfigure[]{
  \includegraphics[angle=0, scale=0.6]{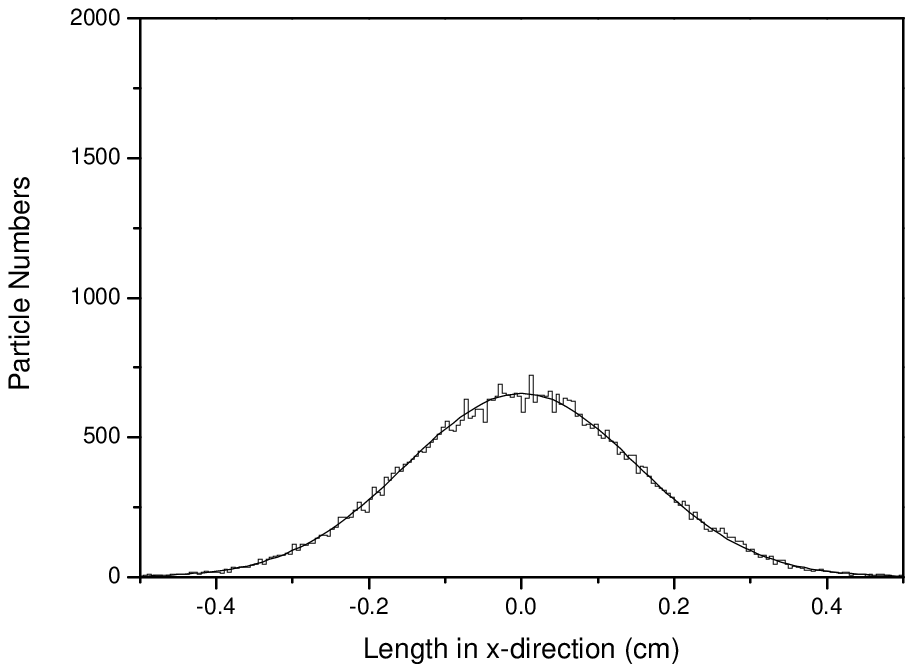}
  \label{fig:horizontal_beam_o}
  }\quad
 \caption{\label{fig:horizontal_beam}
          Particle distributions on the horizontal axis with histograms for
          the PARMILA results and real lines for the model
          calculations: (a) after a focusing quadrupole (b) after a drift space
          (c) after a defocusing quadrupole (d) after a drift space using
          the same normalization as the input beam distribution function.}
 \end{figure}

\begin{figure}[htb]
 \centering
 \hspace{-1cm}
  \subfigure[]{
  \includegraphics[angle=0, scale=0.6]{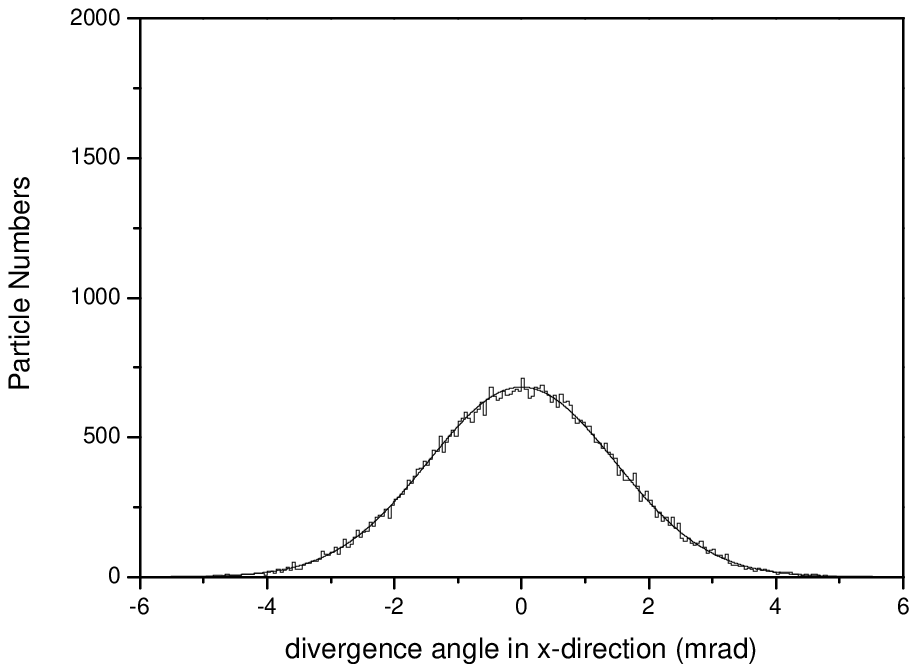}
  \label{fig:horizontal_beam_fp}
  }\quad
  \subfigure[]{
  \includegraphics[angle=0, scale=0.6]{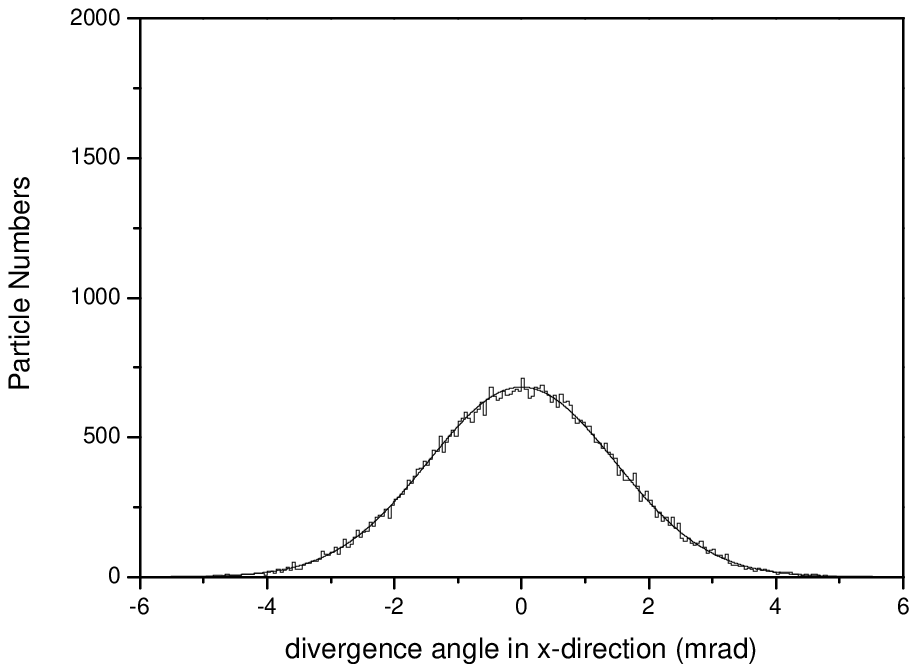}
  \label{fig:horizontal_beam_op}
  }\\
 \hspace{-1cm}
    \subfigure[]{
  \includegraphics[angle=0, scale=0.6]{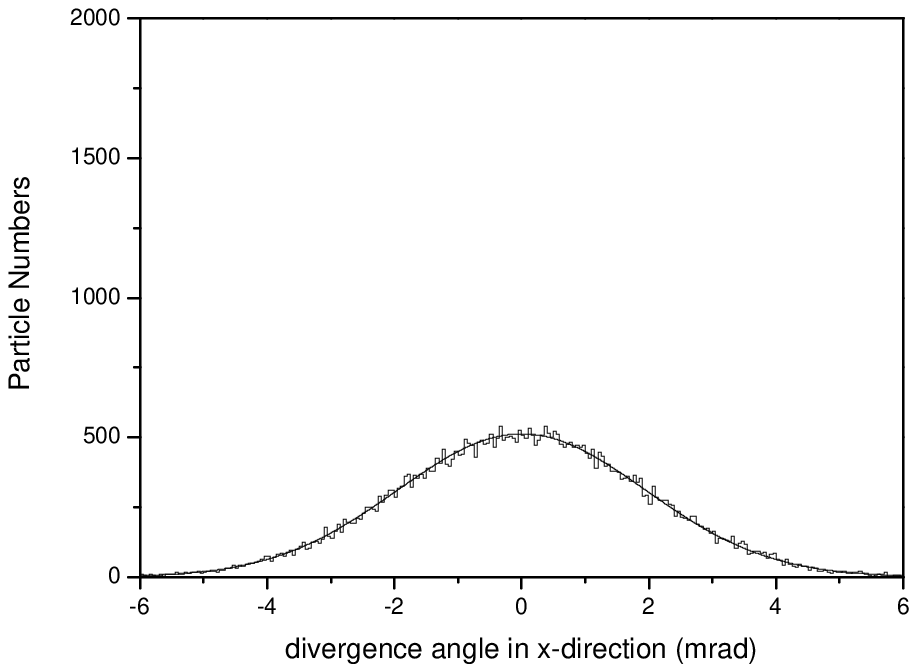}
  \label{fig:horizontal_beam_dp}
  }\quad
  \subfigure[]{
  \includegraphics[angle=0, scale=0.6]{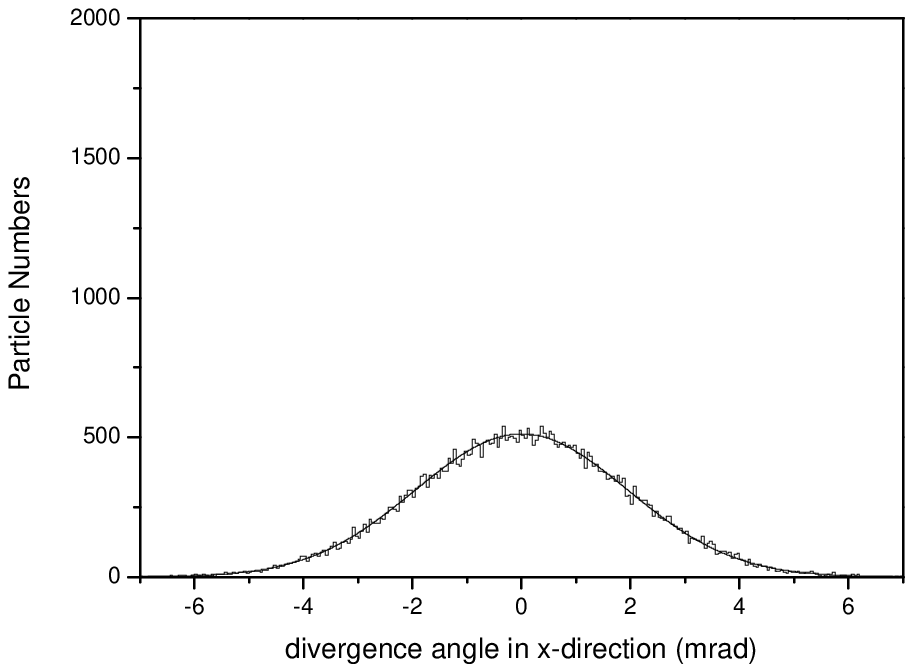}
  \label{fig:horizontal_beam_op}
  }\quad
 \caption{\label{fig:horizontal_beam_p}
          Particle distributions on the horizontal divergence axis 
          with histograms for the PARMILA results and real lines for the model
          calculations using Fourier transformation of the beam wave function
          : (a) after a focusing quadrupole (b) after a drift space
          (c) after a defocusing quadrupole (d) after a drift space using
          the same normalization as the input beam distribution function.}
 \end{figure}

\begin{figure}[htb]
 \centering
 \hspace{-1cm}
  \subfigure[]{
  \includegraphics[angle=0, scale=0.6]{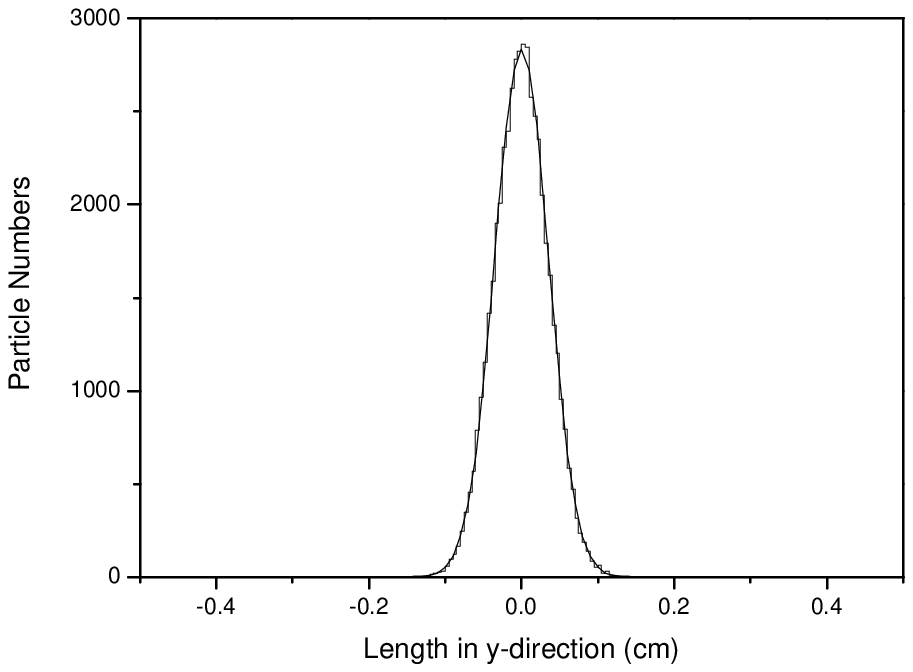}
  \label{fig:vertical_beam_d}
  }\quad
  \subfigure[]{
  \includegraphics[angle=0, scale=0.6]{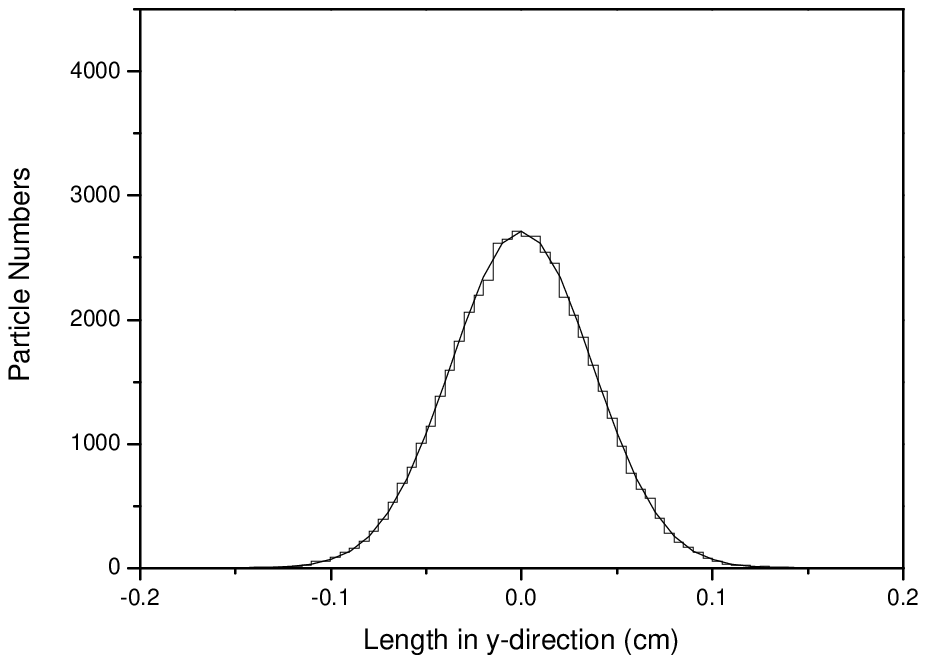}
  \label{fig:vertical_beam_o}
   }\\
 \hspace{-1cm}
    \subfigure[]{
  \includegraphics[angle=0, scale=0.6]{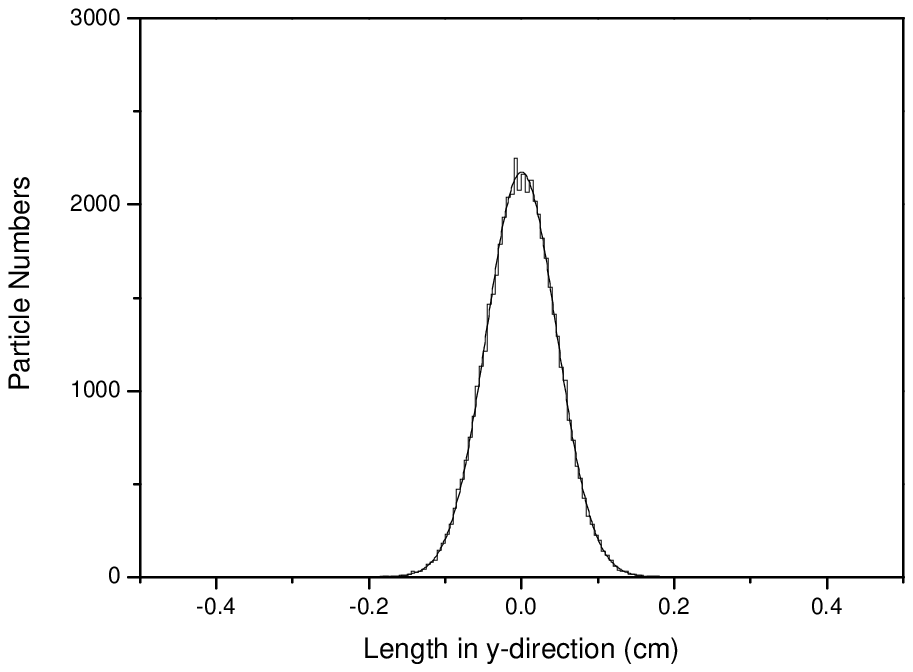}
  \label{fig:vertical_beam_f}
  }\quad
  \subfigure[]{
  \includegraphics[angle=0, scale=0.6]{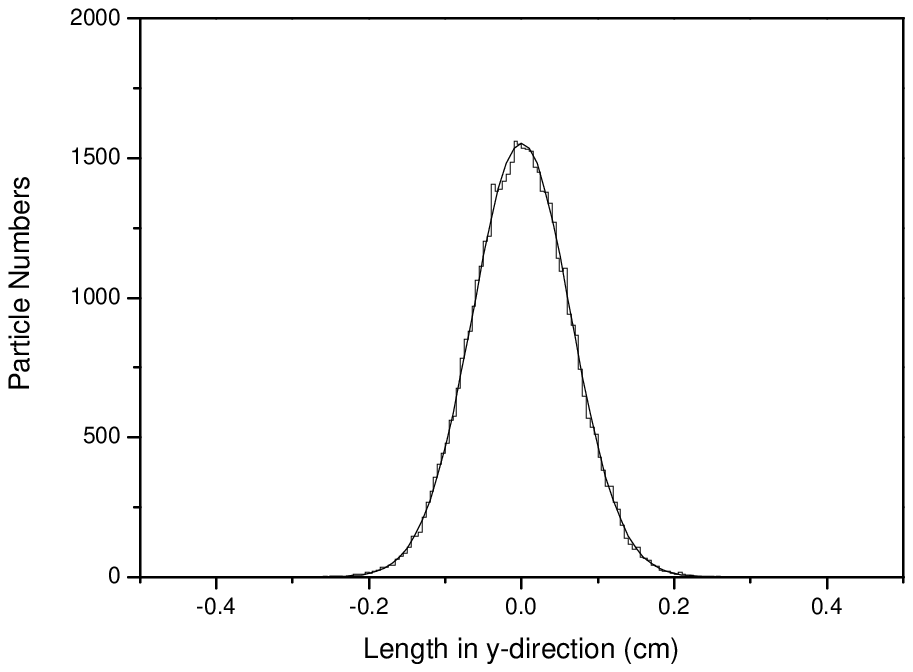}
  \label{fig:vertical_beam_o}
  }\quad
 \caption{\label{fig:vertical_beam}
          Particle distributions on the vertical axis with histograms for
          the PARMILA results and real lines for the model
          calculations: (a) after a defocusing quadrupole (b) after a drift space
          (c) after a focusing quadrupole (d) after a drift space using
          the same normalization as the input beam distribution function.}
 \end{figure}

\begin{figure}[htb]
 \centering
 \hspace{-1cm}
  \subfigure[]{
  \includegraphics[angle=0, scale=0.6]{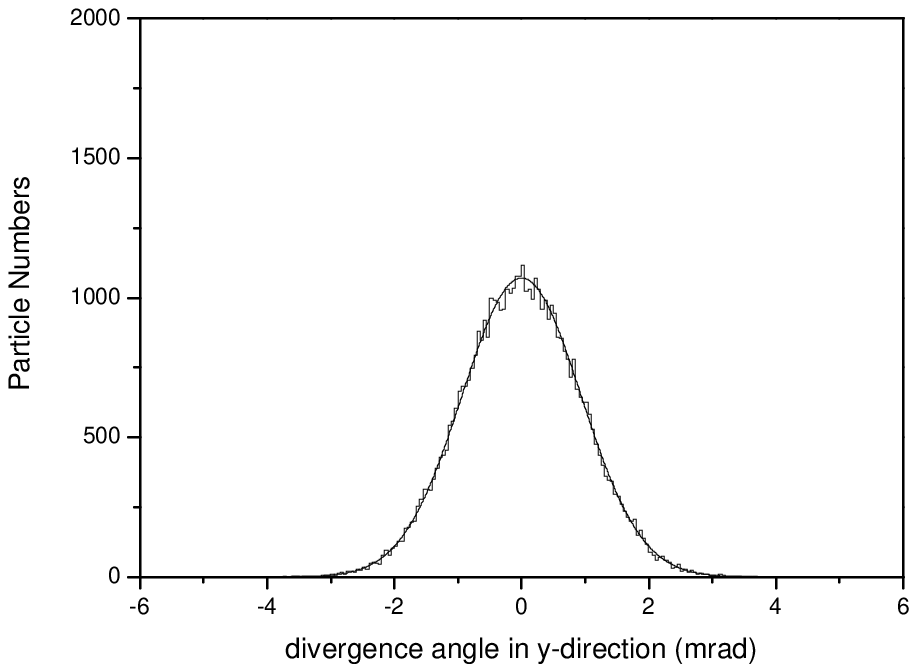}
  \label{fig:vertical_beam_dp}
  }\quad
  \subfigure[]{
  \includegraphics[angle=0, scale=0.6]{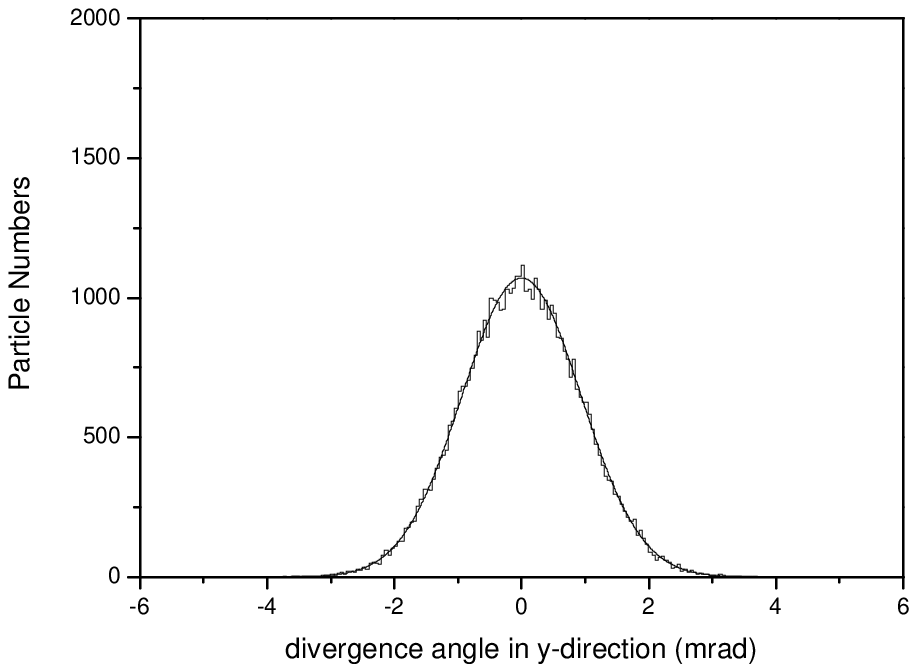}
  \label{fig:vertical_beam_op}
  }\\
 \hspace{-1cm}
    \subfigure[]{
  \includegraphics[angle=0, scale=0.6]{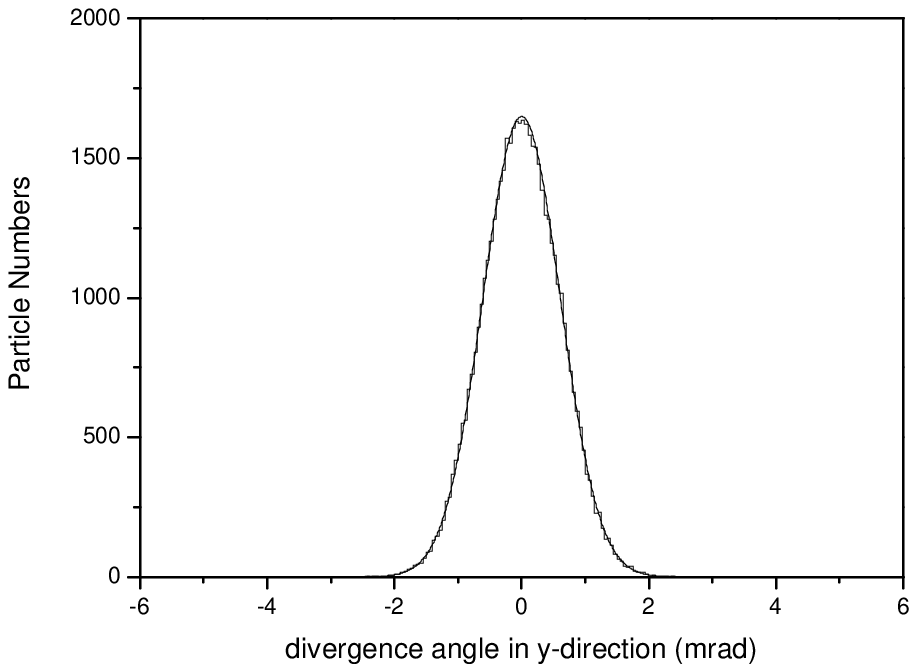}
  \label{fig:vertical_beam_fp}
  }\quad
  \subfigure[]{
  \includegraphics[angle=0, scale=0.6]{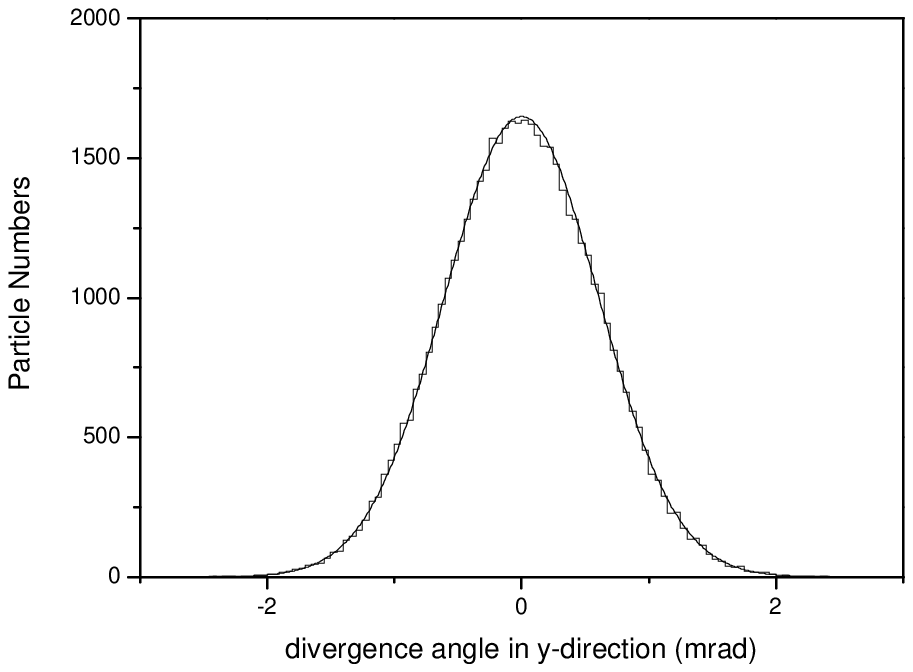}
  \label{fig:vertical_beam_op}
  }\quad
 \caption{\label{fig:vertical_beam_p}
          Particle distributions on the vertical divergence axis 
          with histograms for the PARMILA results and real lines for the model
          calculations using Fourier transformation of the beam wave function
          : (a) after a defocusing quadrupole (b) after a drift space
          (c) after a focusing quadrupole (d) after a drift space using
          the same normalization as the input beam distribution function.}
 \end{figure}

Figure~\ref{fig:horizontal_beam} and Figure~\ref{fig:horizontal_beam_p} show the
particle distributions at the end of each optical element.
The histograms and real lines represent the PARMILA simulation results and 
the model calculations in the $x$ and $x'$ directions, respectively. 
Since the beam wave functions at each stage are Gaussian-type in the $x$ direction,
the wave functions in the $x'$ direction can be obtained by applying
Eq.~(\ref{eq:phi_x}). 
Corresponding figures in the
$y-y'$ space are given in Figure~\ref{fig:vertical_beam} and 
Figure~\ref{fig:vertical_beam_p}. 
We note that the distribution functions are proportional to the
$|\psi_{1x}(x)|^2$ ($|\psi_{1y}(y)|^2$) and $|\phi_{1x}(x')|^2$ 
($|\phi_{1y}(y')|^2$)
in $x$ ($y$) and $x'$ ($y'$) directions, respectively.
From Figure~\ref{fig:horizontal_beam_p} and Figure~\ref{fig:vertical_beam_p},
we can conclude that the Fourier transformation is a valid method to define 
the wave functions in the divergence directions.
The figures show that the one dimensional Schr\"{o}dinger-type equation of 
thermal wave model describes the PARMILA simulation result 
through the FODO (or DOFO) lattice successfully 
in the two dimensional trace ($x-x'$ or $y-y'$) space. 
In order to check on the result quantitatively, we compared the rms beam sizes 
obtained by the model with the values obtained by the best-fit of 
the PARMILA result. 
It is summarized in Table~\ref{table:result}.
It shows that the model results are the same as the simulation to
within 0.8 \%.

In conclusion, we studied the one-dimensional Schr\"{o}dinger-type equation
in the thermal wave model which describes the beam behavior in the $x-x'$ or
$y-y'$ spaces. Some relations were obtained between the model parameters and
physical parameters such as 
the twiss parameters and unnormalized rms emittance.
We used a functional method in order to solve the differential equation 
with the Gaussian input distribution under the condition of
the negligible space charge effects.
The main advantage of this functional method is that we can  
calculate the effects of each beam optical element separately. 
The information of each element is summarized in a kernel. 
The final beam wave function of one optical element is obtained easily 
by the Gaussian integration of the product between the kernel and the
initial Gaussian wave function.
We found that there is a good agreement between the PARMILA simulation and the
model calculation if we neglect the space charge effects.
Even though there are some limits to the application of this 
method, this functional method is a very efficient tool to study the 
transverse beam dynamics in the thermal wave model.

\begin{center}
{\bf ACKNOWLEDGEMENTS}
\end{center}

This work is supported by the 21C Frontier R\&D program in the Ministry of 
Science and Technology of the Korean government.

\vspace{0.5cm}

\newpage
\begin{table}[htb]
\caption{\label{table:input_beam}
         The twiss parameters and unnormalized rms emittances of the input beam 
         in the horizontal and vertical directions.}
\begin{center}
\begin{tabular}{l|ccc}\hline 
               & $\alpha$ & $\beta$ (m/rad) & $\epsilon_{rms}$ ($10^{-7}$ m-rad) 
               \\ \hline
horizontal ($x$) axis& 1.62     & 2.41          & 2.46 \\
vertical ($y$) axis  & 2.87     & 1.16          & 2.57 \\ \hline
\end{tabular}

\end{center}

\end{table}
\begin{table}[htb]
\caption{\label{table:result}
         The rms beam sizes obtained by the model and the best fit of the
         PARMILA result in the horizontal and vertical (numbers in parentheses)
         directions.} 
\begin{center}
\begin{tabular}{l|ccc}\hline 
                   & model (mm)   & PARMILA (mm) & deviation (\%) 
                   \\ \hline
After a F(D) lattice & 0.572(0.353) & 0.575(0.355)    & -0.52(-0.56) \\ 
After a drift space  & 0.251(0.369) & 0.253(0.370)    & -0.79(-0.27) \\ 
After a D(F) lattice & 0.554(0.460) & 0.558(0.461)    & -0.72(-0.22) \\
After a drift space  & 1.521(0.644) & 1.530(0.647)    & -0.59(-0.46) \\ \hline
\end{tabular}

\end{center}

\end{table}

\vspace{1cm}
\newpage

\end{document}